\documentclass[12pt]{article}

\usepackage{scicite}

\usepackage{times}

\usepackage{xcolor}
\usepackage{graphicx}
\usepackage{caption}
\usepackage{amsmath}
\usepackage[modulo,pagewise]{lineno}
\pagecolor{white}

\newenvironment{sciabstract}{%
\begin{quote} \bf}
{\end{quote}}

\title{Observation of Topological Photocurrents in the Chiral Weyl Semimetal RhSi}

\date{}

\begin{document}


\baselineskip24pt

\maketitle
\begin{center}
    Dylan Rees$^{1,2}$, Kaustuv Manna$^3$, Baozhu Lu$^4$, Takahiro Morimoto$^1$, Horst Borrmann$^3$, Claudia Felser$^3$, J. E. Moore$^{1,2}$, Darius H. Torchinsky$^{4*}$, J. Orenstein$^{1,2*}$
\end{center}
$^{1}$Department of Physics, University of California, Berkeley, California 94720, USA\\
$^{2}$Materials Science Division, Lawrence Berkeley National Laboratory Berkeley, California 94720, USA\\
$^{3}$Max Planck Institute for Chemical Physics of Solids, Dresden, D-01187, Germany\\
$^{4}$Department of Physics, Temple University, Philadelphia, Pennsylvania 19122, USA\\
$^\ast$To whom correspondence should be addressed. E-mail: dtorchin@temple.edu (D.H.T.); jworenstein@lbl.gov (J.O.)
\\
\begin{sciabstract}
Weyl semimetals are crystals in which electron bands cross at isolated points in momentum space. Associated with each crossing point (or Weyl node) is an integer topological invariant known as the Berry monopole charge. The discovery of new classes of Weyl materials is driving the search for novel properties that derive directly from the Berry charge. The circular photogalvanic effect (CPGE), whereby circular polarized light generates a current whose direction depends on the helicity of the absorbed photons, is a striking example of a macroscopic property that emerges from Weyl topology. Recently, it was predicted that the rate of current generation associated with optical transitions near a Weyl node is proportional to its monopole charge and independent of material-specific parameters. In Weyl semimetals that retain mirror symmetry this universal photogalvanic current is strongly suppressed by opposing contributions from energy equivalent nodes of opposite charge. However, when all mirror symmetries are broken, as in chiral Weyl systems, nodes with opposite topological charge are no longer degenerate, opening a window of photon energies where the topological CPGE can emerge. In this work we test this theory through measurement of the photon-energy dependence of the CPGE in the chiral Weyl semimetal RhSi. The spectrum is fully consistent with a topological CPGE, as it reveals a response in a low-energy window that closes at 0.65 eV, in quantitative agreement with the theoretically-derived bandstucture. 
\end{sciabstract}
Soon after Dirac discovered his celebrated equation describing a relativistic electron, Weyl pointed out \cite{weyl1929} that a massless particle could have a simpler description because the particle's helicity or handedness is constant, independent of reference frame. Although such Weyl fermions were ruled out as fundamental Standard Model particles after the discovery of neutrino masses, an analogue appears in certain semimetals in which nondegenerate bands cross in momentum space\cite{armitage2018}. These crossing points (or Weyl nodes) act as monopoles of Berry curvature and a theorem by Nielsen and Ninomiya requires the total monopole charge in the Brillouin zone to be zero\cite{nielsen1983}. As a result of this constraint, Weyl nodes cannot be gapped independently and are thus topologically protected.

In recent years, the existence of Weyl nodes and the Fermi arc surface states predicted to accompany them \cite{wan2011} has been conclusively demonstrated by angle-resolved photoemission \cite{xu2015-1, lv2015, xu2015-2, xu2015-3, hasan2017review}. With their existence thus verified, an important goal of future research is to identify the role of Weyl topology in shaping responses to external perturbations. A key step towards this goal is to distinguish topologically-derived responses from those primarily determined by symmetry. The issue arises because topology and symmetry are inextricably linked in Weyl semimetals, as the existence of Weyl nodes requires either inversion or time-reversal symmetry to be broken.

Photogalvanic effects, wherein photocurrents proportional to the light intensity appear in the absence of an applied bias, are examples of responses allowed by symmetry in Weyl semimetals that break inversion. In the circular photogalvanic effect (CPGE), the direction of the current reverses on changing the photon polarization between left and right circular \cite{belinicher1980}. The CPGE has been used effectively to probe broken symmetry states in a variety of condensed matter systems\cite{asnin1979, ivchenko1982, ganichev2003, ganichev2003b}. 

The first hint that topology can shape the CPGE amplitude~\cite{hosur2011} arose in the context of the crossing of non-degenerate bands at the surface of 3D topological insulators such as Bi$_2$Se$_3$. Fig.~1A illustrates how  helicity-dependent photocurrent can arise in such a system as a result of the correlation of the direction of an electron's momentum with that of its spin (or pseudospin). A photon with definite helicity induces a transition that flips the direction of spin, and through spin-momentum locking creates a particle-hole pair that carries a net current. Hosur~\cite{hosur2011} showed that the current associated with photoexcitation of an electron-hole pair at momentum $\mathbf{k}$ was proportional to the Berry curvature, $\Omega(\mathbf{k})$. However, in this two-dimensional (2D) system the net CPGE current vanishes on integration over $\mathbf{k}$ in the presence of n-fold rotational symmetry (for $n\ge3$) . Nonzero CPGE requires lowering the symmetry by in-plane strain, magnetic field, or oblique incidence of the photoexcitation\cite{plank2018}.

Recently de Juan et al.\cite{dejuan2017} showed that, in contrast to the 2D case, rotational symmetry does not cause CPGE to vanish for the 3D bandcrossings that define Weyl semimetals. Instead, the CPGE current from a single Weyl node is proportional to its quantized topological charge and fundamental constants $e$ and $h$. In a non-interacting system this result is independent of material-specific properties and the frequency of the excitation light over a band of wavelengths. The rate of current generation by circularly polarized light is described by the equation,
\begin{equation}
\frac{\textrm{d}j_i}{\textrm{d}t}= i\pi\frac{e^3}{h^2}C\hat {\beta}_{ij}[\mathbf{E}(\omega)\times\mathbf{E}^*(\omega)]_j
\end{equation}
where $\textrm{Tr}{\hat \beta}_{ij}=1$ and $C$ is the monopole charge (or Chern number)\cite{dejuan2017}. In the presence of disorder the photogenerated current will decay with momentum relaxation time $\tau$, yielding a steady state current proportional to $\beta\tau$.

\begin{figure}
\centering
\includegraphics[width=1\textwidth]{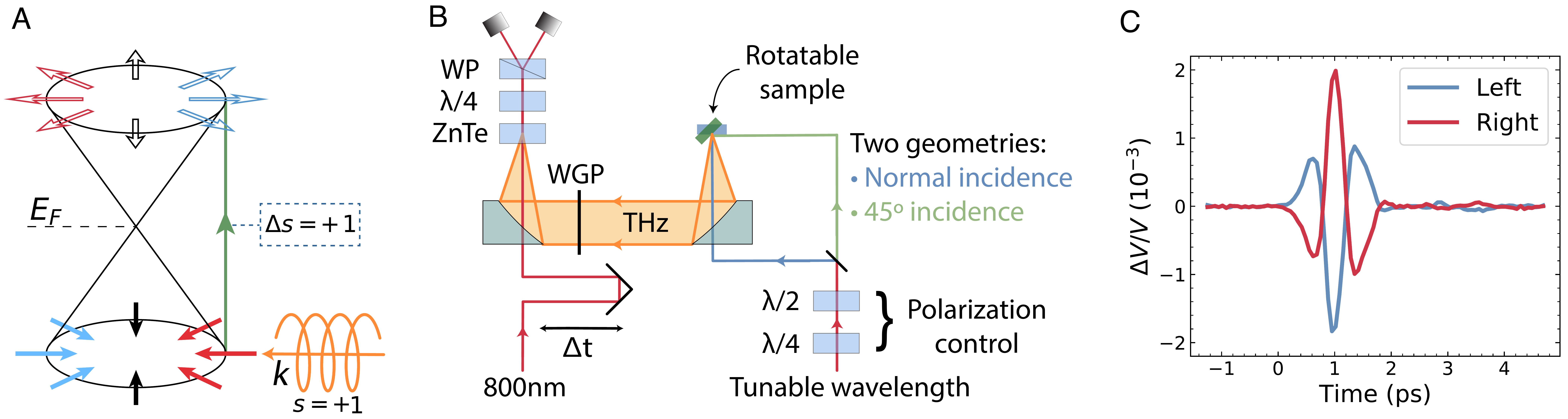}
\caption{ \label{fig:threxp}
\textbf{Photocurrents from Weyl semimetals and experimental apparatus.}
(\textbf{A}) Helical radiation preferentially excites one side of a Weyl cone centered at the Fermi energy, generating a current parallel to the optical wavevector. (\textbf{B}) Schematic of the experimental geometry. Variable wavelength pump light is incident on the sample at either normal or 45$^\circ$ incidence. Terahertz radiation is collected and focused onto a ZnTe crystal for electro-optic sampling. PD, WP, and WGP refer to photodiode, Wollaston prism, and wire grid polarizer, respectively. (\textbf{C}) Individual terahertz pulses measured from left- and right-circularly polarized 2000 nm pump light at 45$^\circ$ angle of incidence. Their difference is the photon-helicity-dependent CPGE signal.}
\end{figure}

Although each Weyl node contributes a quantum of CPGE, this direct signature of topological charge is hidden in systems that retain mirror symmetry, which requires that nodes of opposite charge are degenerate in energy. This leads to an exact cancellation of the CPGE current for pairs of perfectly symmetric Weyl nodes. Despite this, nonzero CPGE is seen in mirror symmetric Weyl semimetals such as TaAs\cite{ma2017,sun2017,sirica2018,gao2019,ji2018,ma2019} as a consequence of departures from symmetric dispersion that occur in real systems, for example curvature or tilting of the Dirac cones\cite{chan2017}. However, in such systems the CPGE amplitude is not a universal topological property uniquely related to the Berry monopole charge.

The properties of chiral Weyl semimetals, in which all mirror symmetries are broken, are qualitatively different from mirror preserving materials such as TaAs \cite{dejuan2017,chang2018}. In chiral structures, isolated Weyl nodes can occur at time-reversal invariant momenta. As a result, they can be separated by wavevectors on the order of the full Brillouin zone, allowing for a richer structure of Fermi arc surface states \cite{bradlyn2016,chang2017,tang2017}. Of more direct relevance to the CPGE, Weyl nodes with opposite topological charge need not be degenerate in energy in chiral media. Thus, it is possible for one node to lie near the Fermi energy, $E_F$, while its oppositely charged partner is below. Transitions near the node below $E_F$ are Pauli blocked at sufficiently low photon energy, and a quantized CPGE (QCPGE) arising from the Weyl node near $E_F$ will emerge.

Chiral semimetals can host multiple bandcrossings with monopole charges $C$ larger than one. Despite higher multiplicity and band curvature in these multifold fermion systems, it was shown theoretically that approximate CPGE quantization continues to hold \cite{chang2017,flicker2018,dejuan2019}. Further, the magnitude of the CPGE is enhanced for multifold compared to Weyl fermions because of the greater topological charge.

RhSi is a structurally chiral material proposed as an ideal candidate to exhibit a QCPGE. The prediction of multifold fermion dispersion and exotic Fermi arcs\cite{chang2017,tang2017} was confirmed recently by ARPES measurements in this compound and in isostructural materials \cite{sanchez2019,takane2019,schroter2019}. The QPGE is predicted to have an especially simple form in this family of compounds because in their cubic space group, P2$_13$ ($\#198$), the dimensionless anisotropy tensor ${\hat\beta}_{ij}$ reduces to the unit tensor multiplied by a scalar $\beta=1/3$. Further, band theory predicts a large energy splitting between the two nodes of opposite charge, such that the regime of Pauli blocking extends to a photon energy of approximately 0.7 eV, well into the near-infrared range\cite{flicker2018}.

A schematic of the apparatus for generation and detection of the CPGE is shown in Fig.~1B. The excitation source was an optical parametric amplifier pumped by an amplified Ti:Sapphire laser, producing wavelength tunable pulses from 1150-2600 nm (0.48-1.1 eV) and duration $\tau_{pulse}\approx$ 100 fs. Photogalvanic current parallel to the surface of the RhSi crystal radiates a THz pulse into free space that is collected and collimated by off-axis parabolic mirrors and then focused onto a ZnTe crystal for time-resolved electro-optic sampling of the THz transient. This all-optical technique avoids artifacts from asymmetric electrical contacts and laser-induced heating, and enables precise determination the direction of the current through measurement of the THz polarization. Fig.~1C shows typical THz transients measured at photon energy 0.60 eV. The reversal of polarity between left and right circular photoexcitation is the defining property of the CPGE. 

Before examining the CPGE spectrum, we first tested that the CPGE and linear PGE (LPGE) currents obey the polarization properties consistent with the space group symmetry of RhSi. Because $\hat{\beta}_{ij}$ is predicted to be diagonal, the CPGE current should obey the relation, $\mathbf{j}\propto\beta(\mathbf{E}\times \mathbf{E}^*)$, and therefore be directed parallel to the wavevector of light, independent of the crystal orientation. The direction of the LPGE current, on the other hand, depends on both the light polarization and the crystal axes. For our measurements, in which the sample was rotated by an angle $\phi$ about the normal to the [111] surface, the direction of the LPGE surface current, $\theta$, is predicted to rotate three times as fast (see Supplementary Information). 

\begin{figure}
\vspace*{-2cm}
\centering
\includegraphics[width=1\textwidth]{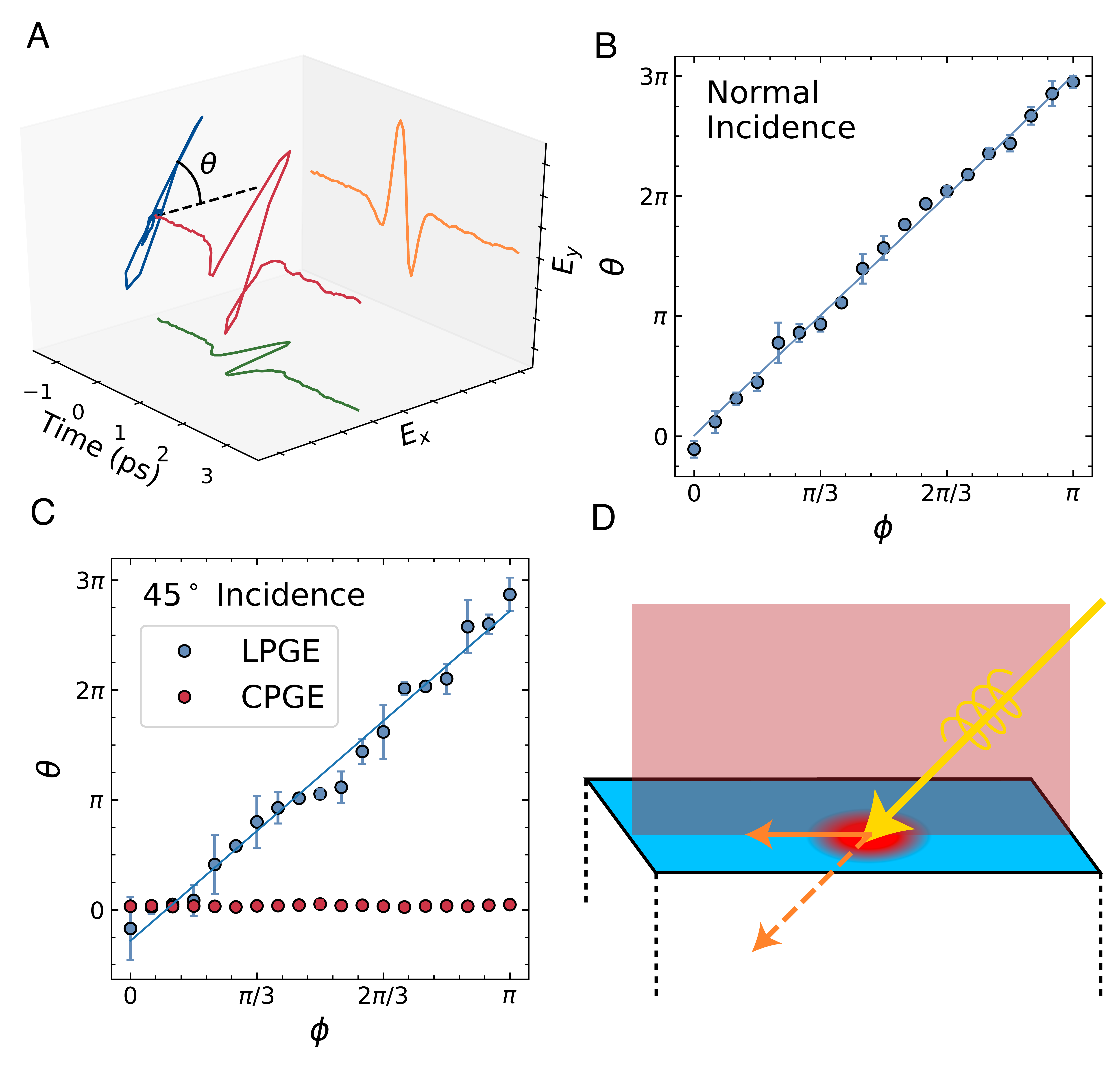}
\caption{\label{fig:exp}
\textbf{ Symmetry of CPGE and LPGE responses in RhSi.}
(\textbf{A}) Measurement of the THz polarization. Orange and green curves show the vertical and horizontal components of the pulse as a function of time. The reconstructed THz pulse (red curve) is then projected onto a plane, showing the direction of linear polarization, $\theta$. (\textbf{B}) Dependence of the angle of LPGE terahertz polarization, $\theta$, on angle of rotation of [111] face about the surface normal, $\phi$, with pump at normal incidence. The relation $\theta=3\phi$ predicted by the space group $P2_13$ symmetry is confirmed. The CPGE signal is below measurement noise level in this geometry. (\textbf{C}) Same as (\textbf{B}) except for 45$^\circ$ incidence. LPGE polarization again varies as $\theta=3\phi$ . CPGE is horizontally polarized independent of the crystal orientation confirming that the CPGE current is parallel to the pump wavevector. (\textbf{D}) Schematic showing that the resulting in plane CPGE current is fixed by the plane of incidence of the pump light. The CPGE current at normal incidence is also normal to the sample, and thus does not radiate into free space. 
}
\end{figure}

To characterize the polarization state of the THz radiation we used a linear polarizer placed before the ZnTe detector to measure both the horizontal and vertical components of the electric field. Fig.~2A shows a typical result of the THz electric field resolved into its two orthogonal components. Fig.~2B shows the direction of the LPGE current as a function of $\phi$ for normal incidence, confirming the relation $\theta=3\phi$. In contrast, the CPGE signal is below measurement noise level at normal incidence, consistent with the prediction that it flows directly into the bulk of the crystal, with zero surface component. At 45$^\circ$ incidence (Fig.~2C), the LPGE current exhibits the same $\theta=3\phi$ dependence and CPGE current is now observed, with direction independent of $\phi$. This latter result is consistent with the expectation that the CPGE current is parallel to the wavevector of the excitation light, because in this case the surface current direction is locked to the plane of incidence (see Fig.~2D), independent of crystal orientation.

Having confirmed that the polarization selection rules are consistent with crystal symmetry, we turn to the dependence of the CPGE amplitude on photon energy $\hbar\omega$ in the range from 0.5 to 1.1 eV. Plotted in Fig.~3 is the $\beta\tau$ product, as our measurements with $\sim$100 fs excitation pulses are in the steady state regime referred to previously. This conclusion is based on the observation that the THz emission waveform does not persist after the laser pulse, indicating that $\tau$ of photogenerated carriers is much shorter than the laser pulse duration. We note further that the relevant $\tau$ is the momentum relaxation time of ``hot'' carriers, which in general is energy dependent and can be shorter than the 8 fs relaxation time determined from equilibrium transport measurements (see Supplementary Information). Converting the measured THz emission to surface current and ultimately $\beta\tau$ requires accounting for multiple wavelength-dependent factors involving the photoexcitation source, the linear optical response of RhSi at the pump laser and THz wavelengths, and the spectral function of the THz detection optics. Propagation of systematic and statistical errors through these multiple factors suggests an order of magnitude uncertainty in the absolute surface current (see Supplementary Information for details). 

The most striking feature of the data is the remarkably sharp and rapid decrease in $\beta\tau$ that occurs when $\hbar\omega$ exceeds 0.65 eV. Above this energy, $\beta\tau$ decays by a factor of $\sim$200 as $\hbar\omega$ reaches 1.1 eV. This spectral feature cannot be accounted for by the aforementioned wavelength-dependent conversion factors, as they vary smoothly through this energy range. Instead we believe the cutoff in $\beta\tau$ is evidence for a crossover in effective Berry charge as a function of $\hbar\omega$, as discussed below. 

\begin{figure}
\centering
\includegraphics[width=1\textwidth]{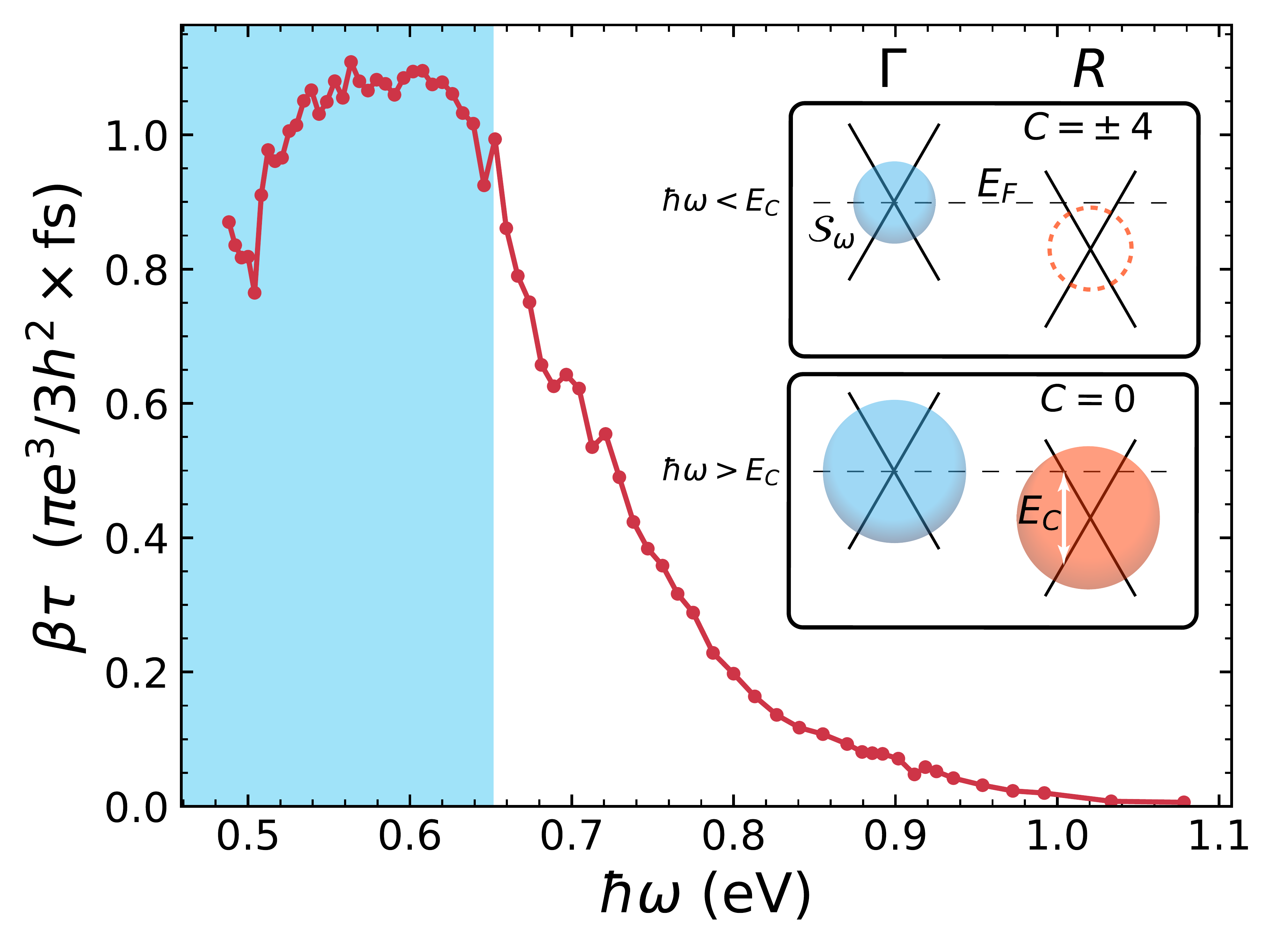}
\caption{\label{fig:symm}
\textbf{CPGE spectrum.}
CPGE amplitude $\beta\tau$ in units of $\frac{\pi e^3}{3h^2}\times$fs as a function photon energy, showing abrupt quenching above 0.65 eV. 
The inset contains a schematic showing the surface $\mathcal{S}_\omega$ in $k$-space defined by the available optical transitions at photon energy $\hbar\omega$. For $\hbar\omega<E_C$, $\mathcal{S}_\omega$ encloses a single node and has integrated Berry flux $C=\pm$4. Above $E_C$ it encloses two topological nodes of opposite chirality and $C=0$. The blue shaded region in the main plot indicates the region where $\mathcal{S}_\omega$ encloses only a single node.
}
\end{figure}

The inset in Fig.~3 illustrates the evolution of the surface $\mathcal{S}_{\omega}$ in $\mathbf{k}$-space defined by the available optical transitions at energy $\hbar\omega$. The CPGE is proportional to the integrated flux of the Berry curvature through $\mathcal{S}_{\omega}$~\cite{dejuan2017}, referred to as $C$. For sufficiently small $\hbar\omega$, $\mathcal{S}_{\omega}$ is a single surface enclosing the $\Gamma$-point and the total Berry flux is equal to the topological charge at $\Gamma$, which is 4. For $\hbar\omega>E_C$ a surface surrounding the $R$-point appears such that $\mathcal{S}_{\omega}$ now encloses two nodes of opposite chirality, driving the net Berry flux, and consequently the CPGE, to zero. Importantly, the measured cut-off energy of 0.65 eV coincides with the value predicted by density functional theory~\cite{chang2017}.

As is clear from Fig.~3, $\beta\tau$ is not frequency independent below 0.65 eV, in apparent disagreement with theory. However, factors specific to RhSi suggest that universal quantized CPGE is not directly observable in this crystal. First, as mentioned above, $\tau$ of hot carriers may depend on photon energy. Second, the prediction of universality depends critically on the assumption that the only allowed optical transitions in the energy range below $E_C$ originate at the $\Gamma$ point. Based on determination of the optical conductivity, $\sigma(\omega)$, we have found that this assumption is not valid.

\begin{figure}\centering
\includegraphics[width=1\textwidth]{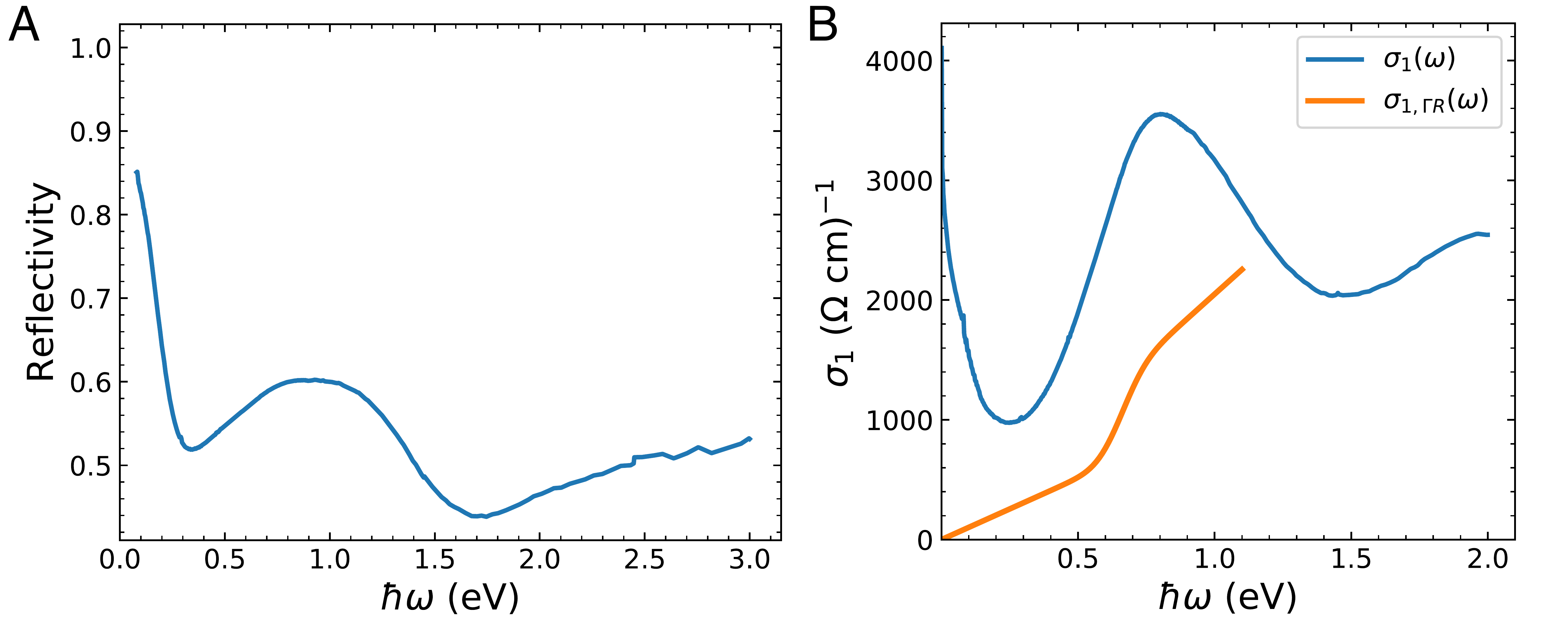}
\caption{\label{fig:spectrum}
\textbf{Reflectivity and optical conductivity.}
(\textbf{A}) Measured reflectivity of RhSi. (\textbf{B}) Optical conductivity determined by reflectivity measurements and Kramers-Kronig analysis (blue curve). The Drude peak is used to infer that the scattering time has value $\tau=8.6$ fs. The orange curve represents the optical conductivity from the $\Gamma$ and $R$ nodes alone~\cite{martinez2019}.
}
\end{figure}

Figs.~4A and 4B show the normal incidence reflectivity and the real part of the optical conductivity obtained by Kramers-Kronig analysis, respectively. The optical conductivity spectrum has a Drude component at low energy and a peak near 0.8 eV that may be associated with transitions near the $M$ point expected from bandstructure calculations\cite{chang2017}. The second curve shows the predicted contribution from Weyl nodes~\cite{martinez2019}, where we estimate disorder and thermal broadening to be 0.1 eV. As is clear from this comparison, only $\sim$$1/4$ of the absorbed photons generate transitions near the $\Gamma$ point that contribute to the topological CPGE. Thus $\beta$ is expected to be modified significantly from the predicted universal value.

The findings reported here lead to new research directions that exploit the chirality of RhSi and related families of compounds. The large energy splitting between Weyl nodes of opposite charge offers a broad window in which the macroscopic response of electrons with definite chirality can be probed. For example, the CPGE photocurrent generated by light at normal incidence flows directly into the bulk of crystal, but decays exponentially with increasing depth. This is an unusual example of a longitudinal current created by a transverse light field. Conservation laws suggest that this CPGE will cause charge and pseudospin to accumulate, which would then couple to longitudinal excitations of the medium such as plasmons and phonons. Thus it may be possible to control the amplitude and phase of these collective modes through the polarization state of incident photons, which would be especially exciting when applied to chiral metals that become superconductors at low temperature.

{\bf References and Notes:}

%
{\bf Acknowledgements:}
We acknowledge D. Parker, Jeremy Johnson and Harold Hwang for useful conversations. {\bf Funding:} J.O., J.E.M. and T.M. were supported by the Quantum Materials program, Director, Office of Science, Office of Basic Energy Sciences, Materials Sciences and Engineering Division, of the U.S. Department of Energy under Contract No. DE-AC02-05CH11231. J.O. received support for optical measurements from the Gordon and Betty Moore Foundation's EPiQS Initiative through Grant GBMF4537 to J.O. at UC Berkeley. J.E.M. received support for travel from the Simons Foundation. D.H.T. acknowledges startup funds from Temple University. K.M., H.B. and C.F. thank the financial support by ERC Advanced Grant No. 742068 ``TOPMAT''. 
{\bf Author Contributions:} 
The experimental setup was initially designed and tested by J.O., D.H.T. and D.R. Spectrally resolved THz emission spectroscopy measurements were performed by D.H.T., D.R., and B.L. Reflectivity measurements and Kramers-Kronig analysis were performed by J.O. and D.R. Partial reflectivity measurements were taken at Advanced Light Source, Beamline 5.4 with the help of Hans Bechtel. Crystal growth, X-ray diffraction, and transport measurements were performed by C.F., K.M., and H.B.
\\
\\
{\bf Supplementary Materials:}
\\
Supplementary Text
\\
Fig S1-S6
\\
References ({\it35-39)}
\clearpage

\section*{Supplementary Text}

\subsection*{Crystal Growth and Structure Refinement}

Single crystals of RhSi were grown from a melt using the vertical Bridgman crystal growth technique. 
Here the crystal growth was performed with an off-stoichiometric composition with slightly excess Si. 
First, a polycrystalline ingot was prepared using the arc melt technique with the stoichiometric mixture 
of Rh and Si metal pieces of 99.99 \% purity. Then the crushed powder was filled in a custom-designed 
sharp-edged alumina tube and finally sealed inside a tantalum tube with argon atmosphere. 
The temperature profile for the crystal growth was controlled with a thermocouple attached at the 
bottom of the tantalum ampoule containing the sample. The sample was heated to 1500$^\circ$C and then slowly cooled to
cold zone with a rate of 0.8 mm/h. Single crystals with average dimension of $\sim$15 mm length and $\sim$6 mm diameter were obtained. 
A picture of the grown crystal is shown in the inset of Fig. S1. The crystals were analyzed with a white beam 
backscattering Laue X-ray diffraction technique at room temperature. The samples show very sharp spots 
that can be indexed by a single pattern, revealing excellent quality of the grown crystals without any twinning or domains. 
A Laue diffraction pattern of the oriented RhSi single crystal superposed with a theoretically 
simulated pattern is presented in Fig. S1. The structural parameters were determined using a 
Rigaku AFC7 four-circle diffractometer with a Saturn 724+ CCD-detector applying graphite-monochromatized Mo-K$\alpha$ radiation. 
The crystal structure was refined to be cubic P2$_1$3 (\#198) with lattice parameter, a=4.6858(9) \r{A}.

\subsection*{Material Symmetries}
\subsubsection*{1. Nonlinear Tensor in the [111] basis}

The second-order optical nonlinearity generates currents at both the sum and difference frequencies of the applied electric field. LPGE and CPGE correspond to the current generated at the difference frequency,
\begin{equation}
j_i\propto \sigma_{ijk}E_jE_k^*
\end{equation}
For cubic space group $P2_13$ the only nonvanishing elements of $\sigma_{ijk}$ have indices $xyz$ and permutations. The elements with even permutations of $xyz$ are equal to $\sigma_{xyz}$ and odd permutations are equal to $\sigma_{xyz}^*$. If we write $\sigma_{xyz} = \alpha+i\gamma$ where $\alpha$ and $\gamma$ are both real, the structure of the third rank tensor can be displayed in the form,

\begin{equation}
\sigma^{(2)} = \left(
\begin{array}{ccc}
 \left(\begin{array}{c}0 \\ 0 \\ 0 \end{array}\right)& \left(\begin{array}{c}0 \\ 0 \\ \alpha+i\gamma\end{array} \right) & \left(\begin{array}{c}0  \\ \alpha-i\gamma \\ 0\end{array} \right)  \\
  \left(\begin{array}{c}0 \\ 0 \\ \alpha-i\gamma \end{array}\right)& \left(\begin{array}{c}0 \\ 0 \\ 0 \end{array} \right) & \left(\begin{array}{c} \alpha+i\gamma  \\ 0 \\ 0\end{array} \right)  \\
  \left(\begin{array}{c}0 \\ \alpha+i\gamma \\ 0 \end{array}\right)& \left(\begin{array}{c} \alpha-i\gamma \\ 0 \\ 0\end{array} \right) & \left(\begin{array}{c}0  \\ 0 \\ 0\end{array} \right)  \\
\end{array}
\right)
\end{equation}
where the element $\sigma_{ijk}$ is the $k$th element of the column vector in the $i$th row and $j$th column of the matrix. 

\subsubsection*{2. Transformation properties of the CPGE}
The circular photogalvanic current can be written in terms of the photon helicity,
\begin{equation}
j_i\propto\beta_{ij}(\textbf{E}\times \textbf{E}^*)_j.
\end{equation}
The second rank CPGE tensor is contracted from the third-rank conductivity tensor according to the relation,
\begin{equation}
\beta_{ij}=\sigma_{ikl}\epsilon_{jkl},
\end{equation}
where $\epsilon_{jkl}$ is the unit antisymmetric tensor. Substitution of the conductivity tensor for the RhSi space group (Eq. 2) yields,
\begin{equation}
\beta_{ij}=i\beta \delta_{ij}, 
\end{equation}
where $\delta_{ij}$ is the Kronecker delta. Substitution into Eq. 3 yields,
\begin{equation}
\textbf{j}\propto i\beta\textbf{E}\times \textbf{E}^*,
\end{equation}
which shows that for the case of space group $P2_13$ the CPGE current is always directed parallel to the helicity vector, regardless of its direction with respect to the crystal axes.

\subsubsection*{3. LPGE sample rotation dependence}
We use Rodrigues' rotation formula to transform this tensor into the basis where $z'$ is parallel to the [111] direction in the crystal basis, which yields
\begin{equation}\label{eq:tensor111}
\sigma^{(2)} \propto\left(
\begin{array}{ccc}
 \left(\begin{array}{c}-\alpha\sqrt{2}\\ 0\\ -1\end{array}\right) & \left(\begin{array}{c}0\\ \alpha\sqrt{2}\\ i\gamma\sqrt{3} \end{array}\right) & \left(\begin{array}{c}-\alpha\\ -i\gamma\sqrt{3} \\ 0\end{array}\right) \\
 \left(\begin{array}{c}0\\ \alpha\sqrt{2}\\ -i\gamma\sqrt{3} \end{array}\right) & \left(\begin{array}{c}\alpha\sqrt{2}\\ 0\\ -\alpha\end{array}\right) & \left(\begin{array}{c}\sqrt{3} \alpha \\ -\alpha\\ 0\end{array}\right) \\
 \left(\begin{array}{c}-\alpha\\ i\gamma\sqrt{3} \\ 0\end{array}\right) & \left(\begin{array}{c}-\sqrt{3} \alpha \\ -\alpha\\ 0\end{array}\right) & \left(\begin{array}{c}0\\ 0\\ 2\alpha\end{array}\right) \\
\end{array}
\right).
\end{equation}
Using this tensor we can calculate the LPGE response for fixed linear pump polarization as the crystal is rotated about the $z'$ (or [111]) axis. The crystal rotation corresponds to the transformation $\sigma_{\alpha\beta\gamma}'=R_{\alpha i}(\phi)R_{\beta j}(\phi)R_{\gamma k}(\phi)\sigma_{ijk}$, where,
\begin{equation}
R(\phi)=\left(
\begin{array}{ccc}
 \cos (\phi ) & \sin (\phi ) & 0
   \\
 -\sin (\phi ) & \cos (\phi ) &
   0 \\
 0 & 0 & 1 \\
\end{array}
\right).
\end{equation}
If, for example, the pump polarization is fixed in the $x'$ direction, then the LPGE current depends only on the tensor elements $\sigma_{xxx}'$ and $\sigma_{yxx}'$ and from Eqs. 7 and 8 we have,
\begin{equation}
\begin{split}
\sigma_{xxx}'= R_{xx}^3\sigma_{xxx} + R_{xx}R_{xy}^2\sigma_{xyy} + R_{xy}R_{xx}R_{xy}\sigma_{yxy} + R_{xy}^2R_{xx}\sigma_{yyx}
\\
\propto\left(-\sqrt{2}\cos^3(\phi)+3\sqrt{2}\cos\phi\sin^2\phi\right)
\\
=-\sqrt2\cos(3\phi)
\end{split}
\end{equation}
and
\begin{equation}
\begin{split}
\sigma_{yxx}' = R_{yx}R_{xx}^2\sigma_{yxy} + R_{yx}R_{xy}^2\sigma_{xyy} + R_{yy}R_{xx}R_{xy}\sigma_{yxy} + R_{yy}R_{xy}R_{xx}\sigma_{yyx}
\\
\propto\left(-3\sqrt{2}\cos^2\phi\sin\phi-\sqrt{2}\sin^3{\phi}\right)
\\
=\sqrt2\sin(3\phi).
\end{split}
\end{equation}

From Eqs. 9 and 10 we obtain the dependence of the LPGE current on crystal rotation angle, 
\begin{equation}
\textbf{j}(\phi)\propto \cos(3\phi)\hat{\textbf{x}}' - \sin(3\phi)\hat{\textbf{y}}'.
\end{equation}
Eq. 11 implies that the angle, $\theta$, of the LPGE current relative to the $x'$ axis is given by $\theta = 3\phi$. This is consistent with what is observed and shown in Fig.~2C and 2D in the main text.

\subsection*{Material Properties}

\subsubsection*{1. Linear optical properties}

We performed reflectivity measurements in the range .08eV - 3eV (main text, Fig.~4A) and used Kramers-Kronig analysis to compute the complex index of refraction $\tilde n=n + i\kappa$. This and the complex dielectric permittivity $\tilde\epsilon=\tilde n^2$ are plotted in Fig~S2. We additionally calculate $\sigma_1= 2n\kappa\epsilon_0\omega$ (main text Fig.~4B and Fig.~S2A) along with $\alpha$, $t_s$, $t_p$ and $\theta_{in}$ (Fig.~S4A-C) which are the absorption coefficient, Fresnel transmission coefficients and the angle of refraction for $45^\circ$ angle of incidence. These are defined by 
\begin{equation}
\begin{aligned} 
\theta_{i n} &=\arcsin \frac{\sin \theta_{i}}{n} \\ 
\alpha &=\frac{2 \omega \kappa}{c} \\ 
t_{s}=& \frac{2 \cos \theta_{i}}{\cos \theta_{i}+\tilde n \cos \theta_{i n}} \\ 
t_{p}=& \frac{2 \cos \theta_{i}}{\tilde n \cos \theta_{i}+\cos \theta_{i n}} 
\end{aligned}
\end{equation}
where $\theta_i=45^\circ$ is the angle of incidence of the pump light.

We compare the predicted optical conductivity of the $\Gamma$ and $R$ nodes in RhSi with our measured $\sigma_1$ in Fig.~S3A and Fig.~4B in the main text. The fraction of the total optical conductivity that represents the predicted linear conductivity is shown in Fig. S3B.

\subsubsection*{2. Scattering time ($\tau$)}

Using the measured value of $\sigma_{dc}$, we infer the equillibrium scattering time to be $\tau=8.6$ fs from our results for $\sigma_1(\omega)$ using the form for the Drude conductivity
\begin{equation}
\sigma_1(\omega)=\frac{\sigma_{dc}}{1+\omega^2\tau^2}.
\end{equation}

\subsubsection*{3. Terahertz index of refraction}
The complex index of refraction at THz frequencies determines the impedance mismatch between RhSi and free space. We can obtain an accurate estimate of $\tilde{n}$ from the dc conductivity. Given the value of $\tau$ determined in the previous section, the THz emission lies in the low frequency limit of the optical conductivity, where $\omega\tau\ll1$ and $\sigma(\omega)\rightarrow \sigma_{dc}$. In this regime, the complex permittivity at low frequency is

\begin{equation}
\tilde{\epsilon}(\omega) =- \frac{\tilde{\sigma}(\omega)}{i\omega}=\frac{i\sigma_{dc}}{\omega}.
\end{equation}
The complex index of refraction is then given as,
\begin{equation}\label{eq:thzindex}
\tilde{n} \equiv n + i\kappa = \sqrt{\frac{\tilde{\epsilon}(\omega)}{\epsilon_0}} = (1+i)\sqrt{\frac{\sigma_{dc}}{2\epsilon_0\omega}}
\end{equation}
which at 1 THz is equal to  $57(1+i)$. Although our Kramers-Kronig analysis begins to become somewhat unreliable for photon energies less than $\sim$10meV, it agrees with our analysis and gives $\tilde n(1\textrm{THz})=51 + 57i$, confirming that Eq.~\ref{eq:thzindex} is correct. When calculating $\beta\tau$ in our analysis, we use $\tilde n(\omega)$ to find the frequency dependent transmission of the THz radiation from RhSi into free space.

\subsubsection*{4. Inferring laser pulse length from emitted THz radiation}

It is not feasible using conventional methods such as autocorrelation to characterize the pulse length $T$ of the laser over the entire wavelength range. Lacking a more precise method, we use the THz time traces to estimate the pulse length at each wavelength, since the laser does not necessarily produce the same pulse width across its available wavelength range. We know that the photocurrent scattering time $\tau$ is much shorter than the pulse length, which means that the instantaneous current follows the applied electric field squared. The effect of the OAP collection filters, described in a later section, which modify the spectrum of the terahertz radiation is to apply a second derivative to the pulse waveform, since for low frequencies the transmission function's leading term is $\omega^2$. By making this approximation we find that the full width half maximum of the terahertz pulse $t_1$ is related to $T$ by the equation
\begin{equation}
t_1 = \sqrt{\frac{1-2W(\sqrt e/4)}{2}}T
\end{equation}
where $W(z)$ is the Lambert W function, or product log. This allows us to calculate an approximate pulse length for each wavelength. The results are presented in Fig.~S4A.

\subsection*{Inferring the CPGE amplitude from the detected electro-optic signal}

In this section we describe the normalization factors needed to convert from signal at the electro-optic detector to the CPGE saturation coefficient $\beta\tau$. This process involves the following three steps.

\begin{enumerate}  
	\item Determine the time-dependent electric field in the sample that arises from a photogenerated surface current which depends on laser parameters (intensity at the sample) and material parameters ($\beta$, $\tau$, Fresnel coefficients, etc.).  
	\item Compute the Fourier transform of the terahertz pulse, the apply two filters to it. The first is the frequency-dependent transmission of the radiation into free space, $1/(\tilde n(\omega) + 1)$. The second is the transfer function of the collection optics, $\mathcal F(\omega)$, that quantifies the fraction of radiation that is collected by the system and transferred to the ZnTe detection crystal.  
	\item Compute the inverse Fourier transform of the resulting spectrum, then convert the resulting time-dependent electric field at the ZnTe surface crystal to signal at the output of the biased photodetector scheme.
\end{enumerate}

In order to obtain the value of $\beta\tau$ of RhSi, we assume $\beta = \beta_0 = \pi e^3/3h^2$ and $\tau=1$ fs for all pump frequencies in this calculation. Then, by dividing the amplitude of the measured signal to the expected signal given all experimental parameters, we yield a value for $\beta\tau$ at each pump frequency in units of ($\beta_0\times$fs).

\subsubsection*{1. Calculation of the radiated field from the sample}

Assume we have some CPGE coefficient $\beta$. For circularly polarized light, CPGE is given by
\begin{equation}
\frac{d j}{d t}=\beta|E|^{2}.
\end{equation}
As laser light travels through a material at angle $\theta_{in}$ relative to the normal direction $z$, its intensity decays according to the (power) attenuation coefficient along the direction of propagation, $r_k=z/\cos\theta_{in}$:
\begin{equation}
|E|^{2}=E_{0}^{2} e_{k}^{-\alpha r_{k}}.
\end{equation}
The sheet current density generation rate is then given by
\begin{equation}
\begin{aligned} \frac{d K}{d t} &=\int_{0}^{\infty} d z \frac{d j}{d t} \sin \theta_{i n} \\ &=\beta E_{0}^{2} \sin \theta_{i n} \int_{0}^{\infty} d z e_{k}^{-\alpha z / \cos \theta_{i n}} \\ &=\beta \frac{1}{2 \alpha} E_{0}^{2} \sin 2 \theta_{i n} \end{aligned}
\end{equation}
and the saturation current density is
\begin{equation}\label{eq:kpar}
K=\beta \tau \frac{1}{2 \alpha} E_{0}^{2} \sin 2 \theta_{i n}.
\end{equation}
The factor of $\sin\theta_{in}$ represents the fraction of the current parallel to the surface, which is what radiates into free space. Since the scattering rate $\tau$ is much less than the pulse length $T$, the current amplitude follows the electric field squared amplitude and will radiate in the THz regime. The radiated electric field amplitude can be found as follows. From Amp\`ere's law we have
\begin{equation}
\begin{array}{c}{E \frac{\tilde{n}}{c}=B=\frac{\mu_{0}}{2} K} \\  \\ {\Longrightarrow E=\frac{\mu_{0} c}{2 \tilde{n}} K}.\end{array}
\end{equation}
Using the Fresnel transmission coefficient $t=2\tilde n / (\tilde n + 1)$, the external radiation is given by 
\begin{equation}\label{eq:e-thz-ext}
E_{e x t}=\frac{\mu_{0} c}{2 \tilde{n}} K t=\frac{Z_0}{\tilde{n}+1} K.
\end{equation}
Using Eq.~\ref{eq:kpar} we get
\begin{equation}
E_{e x t}^{T H z}=\frac{\beta Z_{0} \tau}{2 \alpha(\tilde{n}+1)} E_{0}^{2} \sin 2 \theta_{i n}.
\end{equation}
The frequency dependence of the factor $1/(\tilde n(\omega) + 1)$ is shown in Fig.~S5A.
Now we must express $E_0$ in terms of the measured laser parameters average power $P$, repetition rate $f$, spot size $r_0$ and pulse duration $T$. 
At normal incidence the intensity of the electric field of the pump laser at the surface of the sample is given by
\begin{equation}\label{eq:estim3}
I(r,t)=\frac{c\epsilon_0}{2}E^2(r,t)=\frac{c\epsilon_0}{2}E^2_{\textrm{ext}}e^{-r^2/r_0^2}e^{-t^2/T^2}
\end{equation}
Integrating over space and time yields the relation
\begin{equation}\label{eq:e02-laser}
E_{\textrm{ext}}^2=\frac{2PZ_0}{\pi\sqrt{\pi}fTr_0^2},
\end{equation}
where we include a factor of 1/2 to account for $45^\circ$ angle of incidence. This leaves us with a peak THz electric field of 
\begin{equation}\label{eq:ethz-final}
E_{e x t}^{T H Z}=\frac{Z_{0}^{2} \beta \tau t_{s} t_{p} P \sin 2 \theta_{i n}}{\pi \sqrt{\pi} \alpha f T r_{0}^{2}(\tilde n+1)}
\end{equation}
which radiates into free space and eventually is detected through electrooptic sampling. In order to experimentally determine the spectrum of $\beta\tau$, each of the terms in the above equation must be determined as a function of the pump frequency. As discussed earlier, we calculate $t_s$, $t_p$, $\theta_{i n}=\arcsin 1 /(\sqrt{2} n)$ and $\alpha=2 \kappa \omega / c$ as a function of pump frequency based on spectrally resolved reflectivity measurements and Kramers-Kronig analysis which produces the complex index of refraction. These values are plotted in Fig.~S3C-D. The laser power $P$ is directly measured across the laser's spectral range. We use a concave focal length $F=50$ cm mirror to focus light on the sample, gives a focused spot size of $r_0=\frac{2F\lambda}{\pi d}$ where $d$ is the collimated beam diameter.

\subsubsection*{2. Radiation from the photoexcited region}

The THz radiation emitted by the sample is collected by a $45^\circ$ OAP which collimates the beam. A second OAP then focuses it onto a ZnTe electro-optic sampling (EOS) crystal. In order to calculate the fraction of radiated light collected by the OAP, we start with the formula for the vector potential at location $\mathbf r$ from a current density described by $\mathbf j(\mathbf r, t)$,

\begin{equation}
\mathbf{A}(\mathbf{r}, t)=\frac{\mu_{0}}{4 \pi} \int d^{3} \mathbf{r}^{\prime} \frac{\mathbf{j}\left(\mathbf{r}, t_{r}\right)}{\left|\mathbf{r}-\mathbf{r}^{\prime}\right|}
\end{equation}
where $t_r$ is the retarded time. If we assume for the moment radiation at a specific frequency $\omega$, the current density in our experiment is given by 
\begin{equation}
\mathbf{j}\left(\mathbf{r}, t\right)=J_0 \hat x \delta(z) e^{-x^{2} / 2 r_{0}^{2}} e^{-y^{2} / r_{0}^{2}} e^{-\left(t-x' \sin \theta_{i}/c\right)^2 / T^2}
\end{equation}
where the term $x'\sin\theta_i/c$ in the final exponential represents the phase delay across the photoexcited spot due to off-normal incidence at angle $\theta_i=45^\circ$ (Fig.~S6A). The coordinates used in the calculation are shown in Fig.~S6B.

The retarded time at $\mathbf r$ is given by 
\begin{equation}
\begin{aligned} 
t_{r} &=t-\left|\mathbf{r}-\mathbf{r}^{\prime}\right| / c \\ 
&=t-\frac{1}{c} \sqrt{r^{2}-2 \mathbf{r} \cdot \mathbf{r}^{\prime}+r^{\prime 2}} \\ 
& \approx t-\frac{1}{c} r\left(1-\hat{\mathbf{r}} \cdot \mathbf{r}^{\prime} / r\right) \\ 
&=t-\frac{r}{c}+\frac{\hat{\mathbf{r}} \cdot \mathbf{r}^{\prime}}{c} \\ 
&=t-\frac{r}{c}+\frac{1}{c}\left(\sin \theta \cos \phi x^{\prime}+\sin \theta \sin \phi y^{\prime}\right).
\end{aligned}
\end{equation}
where the OAP is in the far field limit $r'\ll r$.

The Fourier transform of the vector potential is
\begin{equation}
\mathbf{A}(\mathbf{r}, \omega)=\frac{\mu_{0}}{4 \pi r} j_{0} \hat{x} \int d t e^{i \omega t} \int d x^{\prime} d y^{\prime} e^{-\frac{x^{2}}{2 r_{0}^{2}}} e^{-\frac{y^{2}}{r_{0}^{2}}} e^{-u^2/T^2}
\end{equation}
where $u=t_{r}-\frac{x^{\prime} \sin \theta_{i}}{c}=t-\frac{r}{c}+\left(\sin \theta \cos \phi - \sin \theta_{i}\right) \frac{x^{\prime}}{c}+\sin \theta \sin \phi \frac{y^{\prime}}{c}$. This gives
\begin{equation}
\begin{aligned} 
\mathbf{A}(\mathbf{r}, \omega)=\frac{\mu_{0}}{4 \pi r} j_{0} \hat{x} e^{-\omega^2T^2/4} e^{i \omega r / c} & \int d x^{\prime} e^{-\frac{x'^{2}}{2 r_{0}^{2}}} e^{i \omega x^{\prime}\left(\sin \theta \cos \phi- \sin \theta_{i}\right ) / c} \\ 
\times \int & d y^{\prime} e^{-\frac{y'^{2}}{r_{0}^{2}}} e^{i \omega y^{\prime} \sin \theta \sin \phi / c} 
\end{aligned}
\end{equation}
\begin{equation}
=\frac{\mu_{0}}{4\sqrt 2 r} r_{0}^{2}\sqrt\pi j_{0} \hat{x}  e^{-\omega^2T^2/4}  e^{i \omega r / c} e^{-\frac{r_{0}^{2} \omega^{2}}{2 c^{2}}\left(\sin \theta_{i}-\sin \theta \cos \phi\right)^{2}} e^{-\frac{r_{0}^{2} \omega^{2}}{4 c^{2}} \sin ^{2} \theta \sin ^{2} \phi }
\end{equation}
The fraction of the total radiation captured by the OAP is given by 
\begin{equation}
\mathcal F (\omega)=\int_{OAP} d \theta d \phi \sin \theta E(\omega, \theta, \phi) \bigg/ \int_{2 \pi} d \theta d \phi \sin \theta E(\omega, \theta, \phi).
\end{equation}
The second integral is integrated over the upper half-sphere ($0<\theta<\pi/2$) because we only consider the radiation emitted away from the sample, not into it. All $\theta$- and $\phi$-independent factors can be removed from the integrand, so we can use the form
\begin{equation}
E(\theta, \phi) \propto \cos \theta e^{-\frac{r_{0}^{2} \omega^{2}}{2 c^{2}}\left(\sin \theta_{i}-\sin \theta \cos \phi\right)^{2}} e^{-\frac{r_{0}^{2} \omega^{2}}{4 c^{2}} \sin ^{2} \theta \sin ^{2} \phi}
\end{equation}
in the integrand. The integral depends on spot size, $r_0$, which is variable across the spectral range of the pump laser. We calculate this integral numerically for each pump wavelength. $\mathcal F (\omega)$ is plotted for several pump wavelengths in Fig.~S5B as a function of $\nu=\omega/2\pi$. For small wavelengths relative to the excitation spot size, the radiation emits at the specular direction relative to the incoming pump light.

\subsubsection*{3. Electro-optic detection using ZnTe}

The last step in the calibration is the conversion of the electric field at the surface of the ZnTe crystal to the signal at the output of the biased photodetector scheme. Detection of the THz field is performed through electro-optic sampling (EOS) in ZnTe (110). In this technique the THz electric field induces transient birefringence, $\Delta n$, in the ZnTe, which is detected by a co-propagating probe beam at 800 nm. Our analysis is based on the detailed studies of the EOS in technique presented in Refs.~\cite{bakker1998,gallot1999}. For THz frequencies below 3 THz we neglect dispersion in ZnTe and assume a real index $n=2.85$ \cite{nahata1996}.

The transient birefringence generates a polarization rotation in the probe beam. We measure the rotation using an optical bias scheme~\cite{brunner2014,johnson2014} 
that yields a gain factor of 88 as compared with the conventional balanced detector measurement. In the conventional scheme, the fractional change in the balanced output is given by, 
\begin{equation}\label{eq:ZnTedisp}
\frac{\Delta V(\tau)}{V}=\frac{\omega n^3r_{41}}{2c}\int_0^L dz\int_{-\infty}^\infty dt E_{THz}(z,t)I_0(z,t-\tau)
\end{equation}
where $\omega$ is the angular frequency of the probe pulse, $c$ is the speed of light, $L$ is the propagation distance through the crystal, $E_{THz}$ is the THz field strength, $r_{41}=4$ pm/V is the electro-optic coefficient of ZnTe at 800~nm, $n=2.85$ is the index of refraction of ZnTe and 
\begin{equation}
I_0(z,t-\tau)=I_0\exp\{-[z-v_g(t-\tau)]^2/(v_gT_{pr})^2\}
\end{equation}
is the normalized intensity of the 800 nm probe beam with pulse duration, $T_{pr}$, which propagates with group velocity $v_g$. 

THz transients with bandwidth less than 3 THz, Eq. \ref{eq:ZnTedisp} simplifies to,
\begin{equation}
\frac{\Delta V(\tau)}{V}=\frac{\omega n^3r_{41}L}{2c}\int_{-\infty}^{\infty} dt E_{THz}(t)I_0(t-\tau).
\end{equation}
Because the duration of the probe pulse is approximately 35 fs, much less that the time scale of the THz transient, we make the approximation that $I_0(t-\tau) \rightarrow \delta(t-\tau)$ to obtain~\cite{brunner2014}
\begin{equation}\label{eq:dVVfinal}
\frac{\Delta V(\tau)}{V}=\frac{\omega n^3r_{41}L}{2c}E_{THz}(\tau).
\end{equation}
An additional factor of $2/(n+1)$ is needed because the THz field is partially reflected at the surface of the ZnTe. Finally, we substitute $E_{THz}$ with the expression in Eq.~\ref{eq:ethz-final} (after applying the frequency dependent collection filters) and as discussed earlier set $\beta\tau=\beta_0\times1$fs. This gives an expected EOS signal for each pump frequency based on laser parameters, material properties and the experimental geometry, and by comparing the measured value with the expected value we obtain $\beta\tau$ in units of ($\beta_0\times$fs). The results are plotted in the main text in Fig.~3.

\clearpage

\begin{figure}
	\centering
	\includegraphics[width=0.5\textwidth]{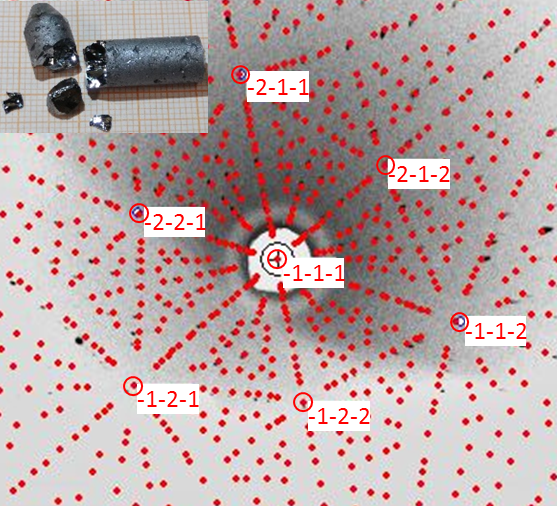}
	\caption*{\label{fig:crystal}
		\textbf{Fig. S1. Crystal growth and diffraction.} Laue diffraction pattern of a [111] oriented RhSi single crystal superposed with a theoretically simulated pattern. Inset shows picture of the grown RhSi single crystal. }
\end{figure}

\clearpage

\begin{figure}
\centering
\includegraphics[width=.7\textwidth]{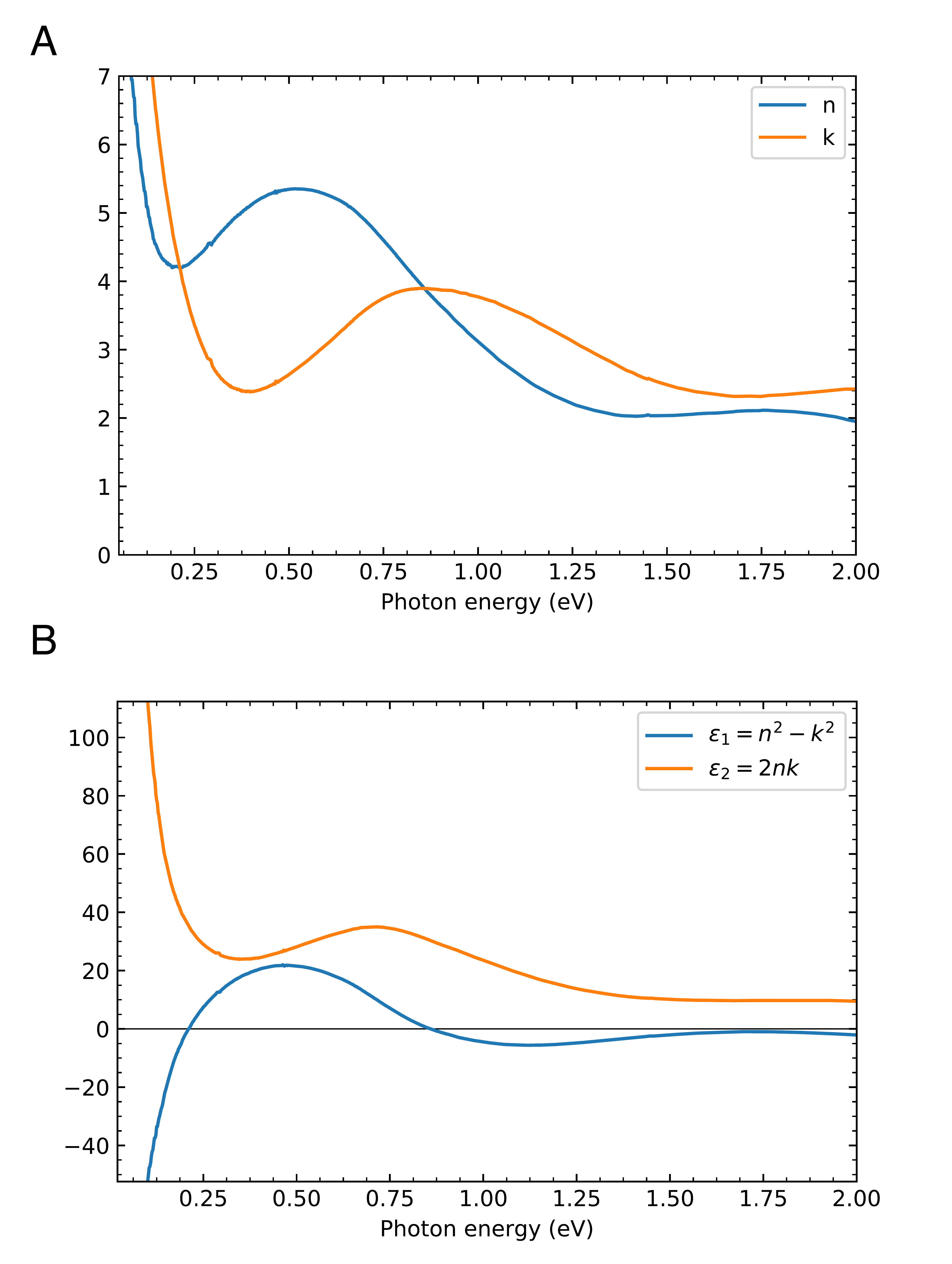}
\caption*{\textbf{Fig. S2. Material properties.} ({\bf A}) The real and imaginary parts of the refractive index.  ({\bf B}) The real and imaginary parts of the complex dielectric function. }
\end{figure}

\clearpage

\begin{figure}
\centering
\includegraphics[width=.7\textwidth]{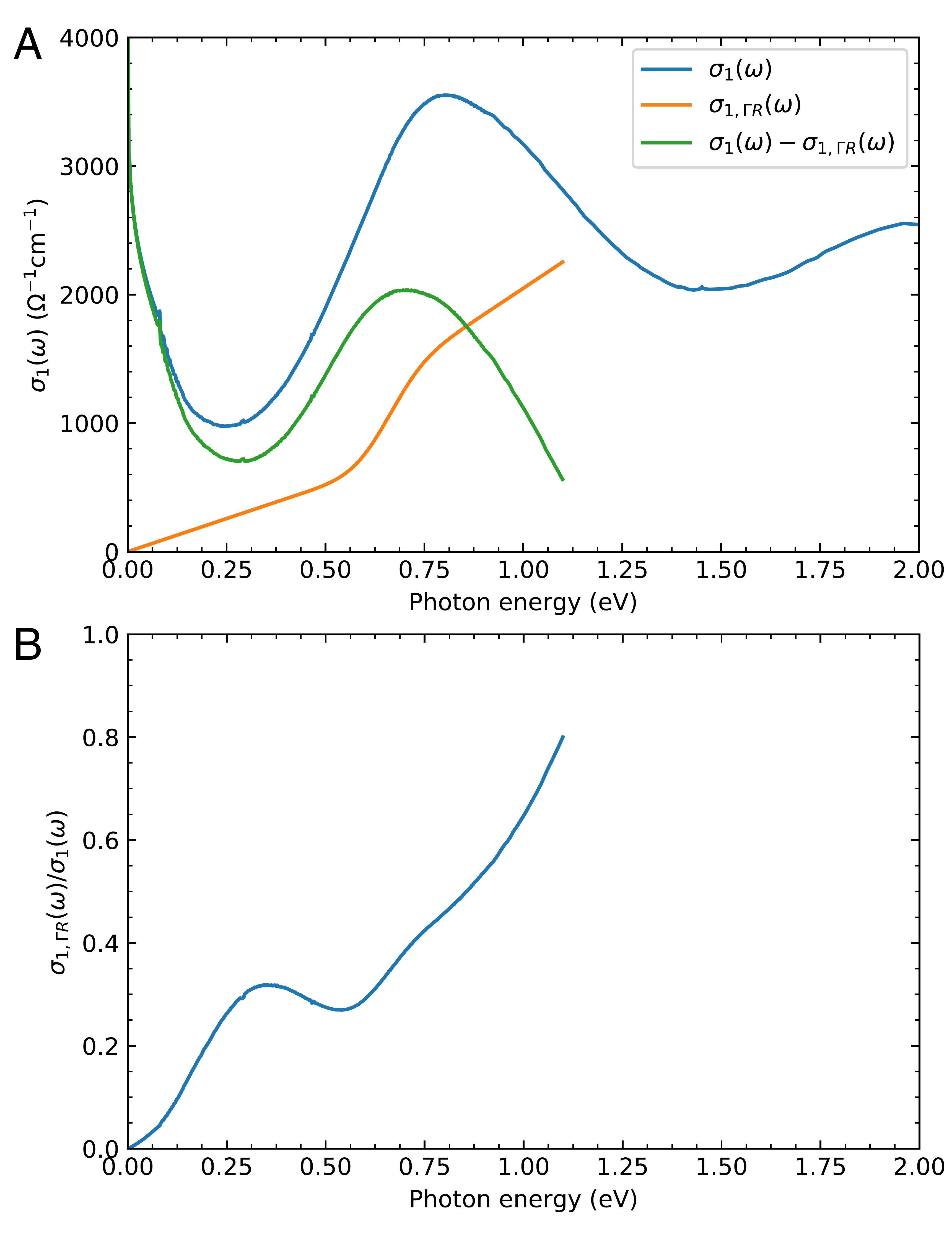}
\caption*{\label{fig:sigma1} \textbf{Fig. S3. Ideal conductivity.} ({\bf A})  Here we compare the ideal $\Gamma$ and $R$ band conductivity, $\sigma_{1,\Gamma R}$, with the total conductivity of  RhSi, $\sigma_1$. ({\bf B}) The fraction of $\sigma_1$ which constitutes the ideal Weyl conductivity.}
\end{figure}

\clearpage

\begin{figure}
\centering
\includegraphics[width=1\textwidth]{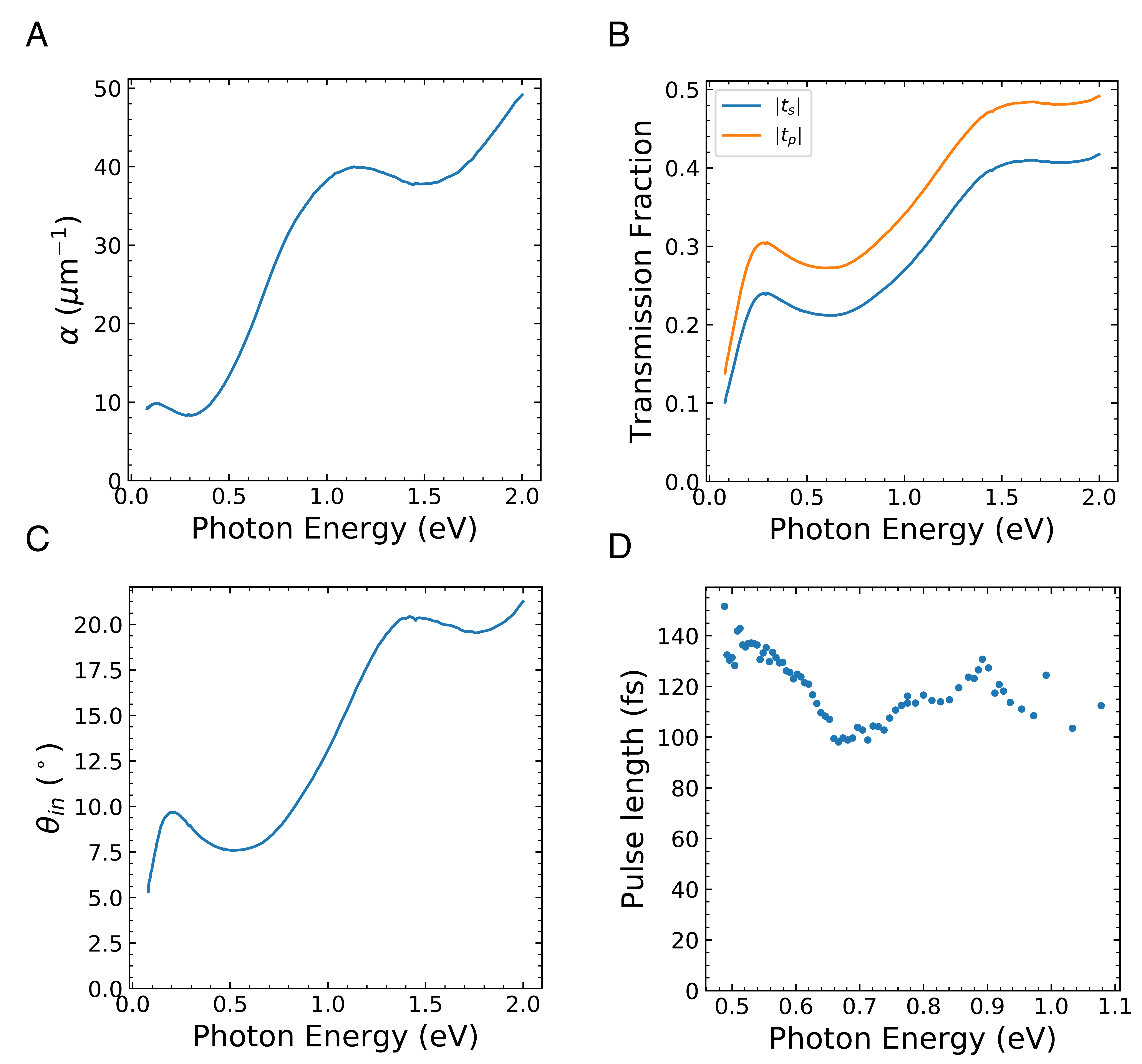}
\caption*{\label{fig:intE} \textbf{Fig. S4. Material properties.} ({\bf A})  Power absorption coefficient $\alpha$.  ({\bf B}) Fresnel transmission coefficient magnitudes $|t_s|$ and $|t_p|$. ({\bf C}) Angle of refraction for incident angle $\theta_i=45^\circ$. ({\bf D}) Pump pulse length $T$ estimated from the emitted terahertz waveforms.}
\end{figure}

\clearpage

\begin{figure}
\centering
\includegraphics[width=1\textwidth]{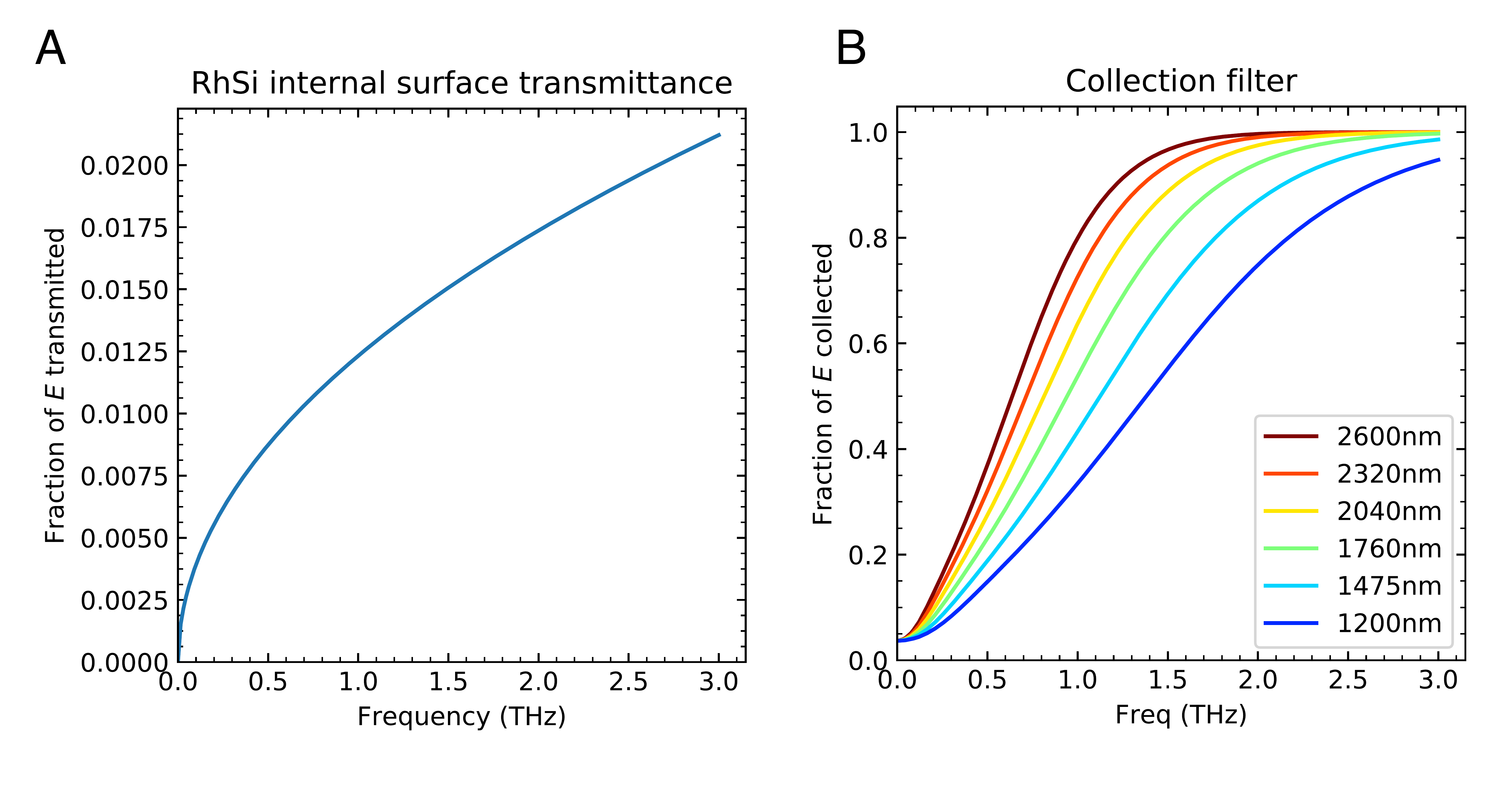}
\caption*{\label{fig:HP_filt}
\textbf{Fig. S5. Surface THz transmission and collection filter.} ({\bf A}) Fraction of terahertz radiation transmitted from bulk into free space, determined by dc conductivity and optical conductivity measurements. ({\bf B}) Fraction of terahertz collected and collimated by the OAP.}
\end{figure}

\clearpage

\begin{figure}
\centering
\includegraphics[width=.8\textwidth]{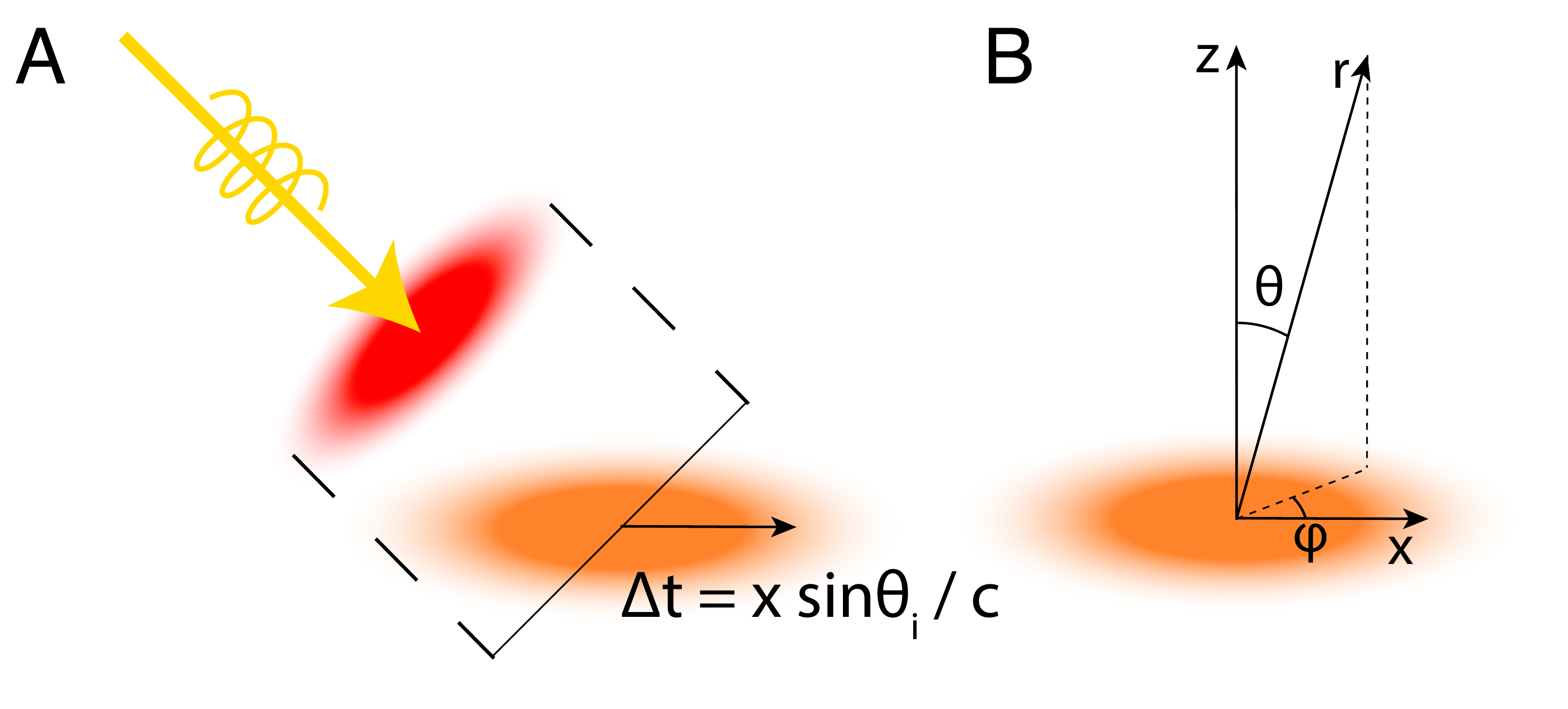}
\caption*{\label{fig:spot}
\textbf{Fig. S6. Filter calculation geometry} ({\bf A}) Illustration of photoexcited current at off-normal incidence. The pump light (red shading) is cast onto the sample which excites a current (orange shading). There is a time delay across the photoexcited region which affects the radiated angle, as in a phased array antenna. ({\bf B}) Illustration of the polar coordinate system used in the filter calculation.}
\end{figure}

\end{document}